# The relation between classical and quantum electrodynamics

## Mario Bacelar Valente

ABSTRACT: Quantum electrodynamics presents intrinsic limitations in the description of physical processes that make it impossible to recover from it the type of description we have in classical electrodynamics. Hence one cannot consider classical electrodynamics as reducing to quantum electrodynamics and being recovered from it by some sort of limiting procedure. Quantum electrodynamics has to be seen not as a more fundamental theory, but as an upgrade of classical electrodynamics, which permits an extension of classical theory to the description of phenomena that, while being related to the conceptual framework of the classical theory, cannot be addressed from the classical theory.

*1. Introduction*

In the wonder world of physics it is usually thought that the relation between the classical and quantum theory is unproblematic and that under a more or less clear procedure we can regard the classical theory as some sort of limit of the quantum theory. Usually these considerations are done in the realm of non-relativistic quantum theory and not much is said about the relation between quantum electrodynamics and classical electrodynamics. Be it the relativistic or non-relativistic theory, we are in the paradoxical situation, which is usually presented as non-paradoxical and natural, that the quantum theory is supposed to contain the classical theory but at the same time needs it for its own foundation. (see e.g. Landau 1974, 13)

In the following I will try to present the idea that classical electrodynamics and quantum electrodynamics form a not fully coherent theoretical framework in which the quantum part has to be seen as an extension of the classical part and not as containing the classical theory. In this way quantum electrodynamics cannot be seen as an independent and more fundamental theory of physics than the classical counterpart.

First, in section 2, I will presented the current classical framework provided by classical electrodynamics and the theory of relativity[1], and address the question of the possible inconsistency of classical electrodynamics and its possible meaning. Then, in section 3, after looking briefly into the development of quantum electrodynamics from the quantization of the classical Maxwell field and the classical Dirac field, I will address the possibility of a classical limit of quantum electrodynamics. Taking into account the limitations of the theory in the description of temporal aspects of physical processes I will defend that properly speaking we cannot reduce classical electrodynamics to quantum electrodynamics.

*2. Classical electrodynamics*

---

[1] I am here taking the theory of relativity to be part of classical physics. This is the view usually adopted, in which, for example, Einstein's theory of gravitation is taken to be a classical field theory (see e.g. Landau and Lifshitz 1971).

Classical electromagnetism as presented nowadays is not the theory developed by J. C. Maxwell. In his more mature work, published in 1873, Maxwell used the Lagrangian formalism to avoid any specific mechanical model of the medium that causes the electric and magnetic phenomena (Harman 1982, 118). His approach was centered on the description of this medium – the ether. The electric current was described as a variation of the polarization – seen as a more fundamental concept – in a material medium (dielectric or conductor); and in this line, the electric charge was considered 'simply' as a spatial discontinuity in the polarization (Darrigol 2003, 164). In practice Maxwell considered the ether and matter as a single medium existing in absolute space, or more exactly, he treated matter "as if it were merely a modification of the ether." (Whittaker 1910, 288; see also Harman 1982, 120) This was a macroscopic theory of the electromagnetic medium that did not make a clear-cut distinction between matter and ether.

In 1892, what can be considered as a new microscopic classical version of electrodynamics was developed by H. A. Lorentz. Adopting the view of microscopic charged particles used in action-at-a-distance theories, Lorentz combined it with the Maxwell theory of the ether in a way that enabled him to explain Fresnel's results regarding the propagation of light in moving bodies. The positive and negative charged material particles would move in the ether without dragging it and only interacting with each other through the mediation of the ether that filled all space: they had a delayed interaction (Whittaker 1910, 420).

Lorentz presented the fundamental equations of his theory as a generalization of the results provided by electromagnetic experiments (Lorentz 1909, 14). This means he made an extension of Maxwell's macroscopic field equations to a microscopic level taking into account his consideration of the charge as a density distribution attached to a microscopic solid body. With this microscopic and atomistic turn, the field equations in Lorentz electrodynamics are given by:

$$\text{div } \mathbf{d} = \rho, \qquad \qquad \text{div } \mathbf{h} = 0,$$

$$\text{rot } \mathbf{h} = 1/c \, (\dot{\mathbf{d}} + \rho \mathbf{v}), \qquad \text{and rot } \mathbf{d} = -1/c \, \dot{\mathbf{h}},$$

where $\mathbf{d}$ is the dielectric displacement, $\mathbf{h}$ is the magnetic force, $\rho$ is the charge density, and $\mathbf{v}$ is the absolute velocity of a microscopic body.

Lorentz considered that the ether pervades all space including the 'interior' of the solid bodies, but being always at rest in relation to absolute space. The law that dictates the influence of the electromagnetic fields, as a manifestation of the internal state of the ether, on the charged bodies can be seen, as in the previous cases, as an extension of the experimental results represented in the force laws of Coulomb and Biot-Savart, and is given by $\mathbf{f} = \mathbf{d} + 1/c \, [\mathbf{v} \times \mathbf{h}]$. These five equations with their underlying assumptions can be considered the core of Lorentz's electrodynamics (McCormmach 1970).

With Lorenz's electrodynamics the conceptual distinction between matter and ether is clearer than in the Maxwell theory. We have a more precise physical characterization of matter, ether, and their interaction, and the scope of application of the theory is extended. In this way we can consider Lorentz's electrodynamics as more fundamental than Maxwell's. But it appears to have a weak spot, which is maintained even after considering A. Einstein's contribution to classical electrodynamics with the downfall of the concepts of ether and absolute space and the rethinking of electrodynamics under the more general theory of relativity.

When in his first works on the subject Lorentz considered the existence of charged corpuscles, he associated them with the ions of electrolysis. P. Zeeman's experimental results that the charge to mass ratio of the particles was one thousand times smaller than supposed indicated that these particles were not the ions of electrolysis. This conceptual distinction led Lorentz to consider the existence of sub-atomic corpuscles (with positive or negative charge), adopting as others the term 'electrons' (Arabatzis 1996, 421-4). Lorentz considered the electron as a charged rigid body, giving to it a "certain degree of substantiality," (Lorentz 1909, 14) to which the laws of motion apply. He modeled the electron as a sphere with a uniformly distributed surface charge. He considered the electron when in motion in relation to the ether (in repose in absolute space) to take the form of an elongated ellipsoid. Considering a very small departure from uniform motion and applying expressions obtained in that case, Lorentz determined "the force on the electron due to its own electromagnetic field." (Lorentz 1909, 38) He found that this effect corresponded to the existence of a mass of electromagnetic origin and was taken to the idea of an effective mass composed of the mechanical mass and the electromagnetic mass. Due to W. Kaufmann's experiments, and considering the mechanical mass not from the point of view of the not yet developed theory of relativity but from Newtonian mechanics, Lorentz even considered the possibility that the electron's mass was all of electromagnetic origin. Lorentz's work was critically examined by H. Poincaré who concluded that a non-electromagnetic internal pressure was needed, so that the electron was stable under the electrostatic repulsion between its elements of charge (Poincaré 1905).

From the point of view of the theory of relativity it is clear that the mass of the electron cannot be solely of electromagnetic origin. The electron's momentum and energy originated by its own field do not form a four-vector. In relativistic mechanics we can consider a particle to be defined by having a determined energy-momentum four-vector (see e.g. Jammer 1961, 164). This definition can be justified without taking into consideration any aspect of electrodynamics, as G. N. Lewis and R. C. Tolman have done: we can determine the relativistic expression for the particle's mass by considering the collision between the particles and postulating a conservation law of momentum and using the relativistic law of addition of velocities (Pauli 1958, 118). From this it is immediate to see that the momentum and the energy of the particle behave under Lorentz transformation as the components of a four-vector. This result can be checked experimentally again without any explicit use of electrodynamics, as was done in the early 1930s by F. C. Champion, who studied the scattering of β-particles with electrons at rest in a Wilson chamber (Zhang 1997, 234). This means that the experimental and conceptual framework of relativistic mechanics can be developed and verified on its own, detached from any electrodynamical considerations. This point is crucial in the analysis of the difficulties that relate to classical models of the electron.

When considering a classical point-like model of the electron, the self-energy is infinite. In 1938, P. Dirac proposed a clear covariant procedure to separate the finite and infinite contributions to the self-energy (Dirac 1938). The infinite contribution to the electron's mass is taken care of by a renormalization procedure in which the observed mass encloses the mechanical mass and the (infinite) electromagnetic mass. The finite effect is a reaction force depending on the derivative of the acceleration. So, when considering the electron's self-energy we have a departure from the Lorentz force equation and obtain an equation – the Lorentz-Dirac equation – in which besides the external force we have present the radiation reaction from the electron's field. This equation has very unphysical solutions. In the absence of any external force the equation admits solutions where the radiation reaction provides a self-acceleration to the charged particle. Choosing appropriate asymptotic conditions this type of solution is avoidable. Nevertheless a problem still remains. When considering the case of a particle subject to an external force, the motion of the particle is affected by the force even before the action of the force: we have a pre-acceleration of the electron before the action of the external force (Barut 1964, 198). It seems that the point-like electron is not a classically acceptable model. However, the pre-acceleration solution, where we have a non-zero acceleration before the external force is applied, appears to be avoidable with a classical extended electron model.

Let us consider a model of the electron consisting in "a charge $e$ uniformly distributed on the surface of an insulator which remains spherical with constant radius $a$ in its proper inertial frame of reference." (Yaghjian 1992, 31) Taking into account the finite velocity of propagation of an electromagnetic disturbance across the 'electron,' in the derivation of the solution of the equation of motion of this 'electron,' no pre-acceleration appears in the solution.

This result has been challenged, and it might be the case that even this model does not resolve the problem of pre-acceleration (Frisch 2005, 62).[2] In this way there seems to be no conceptually unproblematic way to overcome the inconsistency we have in the actual applications of classical electrodynamics where, when considering a particle-field system, the Lorentz laws are applied taking into account energy-momentum conservation while ignoring the self-field of the particle. According to M. Frisch, "the standard way of modeling phenomena involving the interaction between discrete charged particles and electromagnetic fields relies on inconsistent assumptions ... the equation of motion for discrete charges that is used in all applications of classical electrodynamics, which ignores the self-fields of the charge, is inconsistent with the Maxwell equations and the standard principle of energy momentum conservation" (Frisch 2007, 2). Also, in Frisch's view, "there are a host of conceptual problems that arise when one tries to develop a fully coherent and complete classical theory of charged particles interacting with electromagnetic fields—a theory that does not simply ignore self-interaction effects" (Frisch 2007, 3-4). In this way the conceptual problems force upon us an inconsistent approach.

I think that, more than some sort of inconsistency, we are facing here interesting and revealing aspects of classical electrodynamics. One thing that we can conclude from the analysis that led to the inconsistency claim by Frisch, is that we have a limited description of matter within the classical theory. As has been noted regarding classical electrodynamics, "the main problem with taking this theory to be the fundamental theory of the interaction of classical charges and fields is that it is in an important sense incomplete. Without substantive additional assumptions concerning how charged particles are to be modeled, the theory cannot be understood as describing the behaviour of the particle-field system." (Frisch 2005, 47) Basically we only have general rules from relativistic mechanics that give an overall prescription about what general laws matter must 'obey', like the definition of the concept of particle by considering that it must have a certain energy-momentum four-vector, which is independent of any particular model of the particle and the possible inconsistency of any derived force law (as mentioned earlier, the experimental and conceptual framework of relativistic mechanics can be developed and verified on its own, detached from any electrodynamics considerations). We really do not have an elaborated theory of matter. From this I would say that the classical theory is incomplete, in the sense that part of its conceptual framework – the one related with the description of matter – shows severe limitations. But, contrary to the view expressed by Frisch (2007), I find it difficult to regard (in some sense) as inconsistent a theory that we have strong reason to consider incomplete.[3]

---

[2] It is important to notice that this is not a settled matter. For example F. Rohrlich claims that "for a charged sphere there now exist equations of motion both relativistically and nonrelativistically that make sense and that are free of the problems that have plagued the theory for most of this century; these equations have no unphysical solution, no runaways, and no preaccelerations" (Rohrlich 1997, 1054-5; see also Rohrlich 2008)

*3. The relation between classical and quantum electrodynamics*

In this part of the paper I will defend the idea that quantum electrodynamics cannot be seen as a more fundamental theory than its classical counterpart, in the sense that we could recover the classical theory from some sort of limit of the quantum theory. Instead I will propose to see the quantization procedure as a literal 'upgrading' procedure in which we built the quantum part from the classical one in a way that the quantum part is dependent of the underlying classical structure.[4]

I will start, in the first subsection, by considering the quantization procedure. I will then consider briefly the literature addressing the so-called classical limit in quantum electrodynamics. While in the case of quantum mechanics there is a vast literature about this subject, it seems that physics and philosophy have not taken too much into account the question of the relation between classical electrodynamics and quantum electrodynamics. What we find is basically the usual idea of the Planck's constant 'limit' ($h \to 0$) transposed to the case of quantum electrodynamics.

In the second subsection I present part of the argument against the previous simplified view of the relation between the two theories. As is well known the application of quantum electrodynamics to the description of bound-state or scattering problems is made by resort to perturbative approaches (in particular the S-matrix approach). The point is that in the perturbative approach we are calculating integrals that go from $t = -\infty$ to $t = +\infty$. This overall temporal description makes it impossible to associate a particular time interval (duration) to the physical processes described within the theory.[5] The problem is then seeing how the duration of physical processes we have at the macroscopic level (as described by classical theories) could emerge from the quantum realm since the theory is unable to provide any indication of a temporal duration of the processes that are supposed to be occurring at the microscopic level and constituting the macroscopic level where our perception of time is supposed to be taking place.

---

[3] Another aspect, related to the previous one, is that the theory was designed by considering two clearly distinct entities: the field and the particles, and in the usual applications of the theory "electric charges are treated *either* as being affected by fields *or* as sources of fields, but not both" (Frisch 2004, 529). In trying to overcome this approximate approach, the development of the theory faces clear difficulties that within the classical realm seem to have no easy solution (Frisch 2007, 11). In this way, it seems that we might be facing an intrinsically approximate feature in the description of the interaction of radiation and matter at the classical level. If it turns out to be so, we are really facing more than a problem of 'incompleteness' in the description of matter in the classical theory. According to Bohr's views this is also the situation in quantum electrodynamics (Bohr 1932a, 378; Bohr 1932b, 66; see also Rueger 1992, 317-8). That we might be facing similar problems in the two theories can be expected according to the view being proposed here on the relation between the two 'theories'; another example is the renormalization of the electron's mass in the two 'theories'; see e.g. Barut (1964, 190-1) and Schweber (1961, 524-30). However to go into details on these subjects would go beyond the scope of this work, which will be centered on the temporal description of physical processes. In this way the possible inconsistency of the applications of classical electrodynamics when excluding self-interactions (or other seemingly insurmountable problems when including them) will not be of importance to the central ideas being presented here. I include them for 'completeness' in the presentation of classical electrodynamics.

[4] In this work, I am not taking into account different interpretations of quantum theory and different approaches to the quantum theory of measurement (in particular decoherence). I am solely developing my argument within an ensemble interpretation (e.g. Isham 1995, 80-1) that gives a natural connection between the calculations done using quantum electrodynamics and the results from the experimental procedures followed (e.g. Falkenburg 2007, 106 and 207).

Now if we take classicality as emerging from the 'quantum world', the physical processes that at a macroscopic level we see taking place during finite time intervals should emerge from some, even if 'diffuse', assignment at the quantum level of a 'duration' to the physical processes ('constituting' the macroscopic ones). As mentioned, at the level of the usual applications of quantum electrodynamics this is not possible to do: we just have an overall temporal description. In subsection 3 I will consider whether it might be possible within quantum electrodynamics to recover a more classical type of association of a time interval to physical processes. To this purpose I will consider Fierz's (1950) take on Fermi's two-atom system (Fermi 1932). It turns out that the description, using in part the formalism of quantum electrodynamics, of the interaction between two bound electrons cannot be seen as a direct application (or consequence) of the theory. It is more like a piecemeal model that can be seen to include a classical input. It is exactly this classical input (or quantum input at the correspondence level where by design the quantum theory provides results equivalent to those of the classical theory) that makes it possible to associate a temporal time interval – duration – to the process of emission and subsequent absorption of a photon between two bound electrons. In this way it seems that the type of temporal description of physical processes we have at the classical level cannot be seen as emerging from the quantum level of description.

*3.1 The quantization procedure and the so-called classical limit*

Within quantum electrodynamics the starting point are classical fields like the Maxwell field and the Dirac spinor field defined on a Minkowski space-time. We can see the quantization scheme as a set of physical rules that enable an extension of the applicability of the classical concepts to phenomena that while categorized as related to matter, radiation, and their interaction are beyond the classical sphere of description.

Considering the usual quantization procedure, in the case of a free Maxwell field the vector potential can be expanded as

$$A^\mu(x) = \sum_{r\mathbf{k}} \left(\hbar c^2 / 2V\omega_\mathbf{k}\right)^{1/2} \left(\varepsilon_r^\mu(\mathbf{k}) a_r(\mathbf{k}) e^{-ikx} + \varepsilon_r^\mu(\mathbf{k}) a_r^\dagger(\mathbf{k}) e^{ikx}\right).[6]$$

In order that $A^\mu(x)$ can be related to the Maxwell equations a subsidiary condition is imposed, the so-called Lorentz subsidiary condition $\partial_\mu A^\mu(x) = 0$, which at a quantum level has to be changed to a condition on admissible state vectors: $(\partial_\mu A^\mu)\Psi = 0$. In this case the connection between the classical equations and concepts and their quantum upgrades is very direct. Under the quantization scheme $A^\mu(x)$ is now a field operator, and the Fourier expansion coefficients $a_r(\mathbf{k})$ and $a_r^\dagger(\mathbf{k})$ are now, as operators, conditioned by the commutation relations $[A^\mu(\mathbf{x}, t), A^\nu(\mathbf{x}', t)] = 0$, $[\dot{A}^\mu(\mathbf{x}, t), \dot{A}^\nu(\mathbf{x}', t)] = 0$, and $[A^\mu(\mathbf{x}, t), \dot{A}^\nu(\mathbf{x}', t)] = -i\hbar c^2 g^{\mu\nu} \delta(\mathbf{x} - \mathbf{x}')$.

---

[5] In quantum electrodynamics interaction processes are described as exchange of virtual quanta, and it is impossible to associate to the exchange of virtual quanta finite time intervals in opposition to the case of real quanta where this is possible (see main text).
[6] For details about this expression and the quantization of the electromagnetic field see e.g. the textbook by Mandl and Shaw (1984).

In the case of the Dirac equation, it might seem that the situation is not so simple. Initially, the equation was developed by Dirac as a relativistic one-electron wave equation. Soon afterwards, due to the difficulties with the negative-energy solutions, Dirac adopted the so-called hole theory in which the equation became a many-particle equation describing not only electrons but also positrons. By adopting the second-quantization formalism (more exactly P. Jordan's interpretation of it) a quantum field perspective can be given of Dirac's equation (see e.g. Darrigol 1986). In this case, we can consider Dirac's equation as a 'classical' equation of an electron-wave, whose properties can be explored in experiments like the diffraction experiment of Davisson and Germer.[7] Following Jordan, we can consider the quantization of this classical spinor field using in this case anticommutation relations, and obtaining, by a procedure even more simple than the quantization of Maxwell field, the Dirac field operators

$$\psi(x) = \sum_{rp} (mc^2/VE_\mathbf{p})^{1/2} \left( c_r(\mathbf{p}) u_r(\mathbf{p}) e^{-ipx/\hbar} + d_r^\dagger(\mathbf{p}) v_r(\mathbf{p}) e^{ipx/\hbar} \right),$$

$$\overline{\psi}(x) = \sum_{rp} (mc^2/VE_\mathbf{p})^{1/2} \left( d_r(\mathbf{p}) \overline{v}_r(\mathbf{p}) e^{-ipx/\hbar} + c_r^\dagger(\mathbf{p}) \overline{u}_r(\mathbf{p}) e^{ipx/\hbar} \right),$$

where $c_r^\dagger$, $d_r^\dagger$, $c_r$ and $d_r$ obey the anticommutation relations $[c_r(\mathbf{p}), c_r^\dagger(\mathbf{p}')]_+ = [d_r(\mathbf{p}), d_r^\dagger(\mathbf{p}')]_+ = \delta_{rs} \delta_{\mathbf{pp}'}$, with all other anticommutation relations vanishing (Mandl and Shaw 1984, 6).

We can see Dirac's equation as a classical-level description of matter from a wave perspective, but in agreement with the laws of relativistic mechanics. In particular, the relativistic Hamiltonian $H = c(m^2c^2 + p_1^2 + p_2^2 + p_3^2)^{1/2}$ can be seen as fundamental in the derivation of Dirac's equation (see e.g. Dirac 1958, 255). In his 1928 paper on the relativistic one-electron wave equation, Dirac considered that the relativistic wave equation of the electron should be linear in $p_0 = i\hbar \partial/(c\partial t)$ and $p_r = -i\hbar \partial/(c\partial x_r)$ with $r = 1, 2, 3$; in this way it had the form $(p_0 + \alpha_1 p_1 + \alpha_2 p_2 + \alpha_3 p_3 + \beta)\psi = 0$. The matrices $\alpha_1$, $\alpha_2$, $\alpha_3$ and $\beta$, are determined by the relativistically invariant equation $(p_0^2 - m^2c^2 - p_1^2 - p_2^2 - p_3^2)\psi = 0$ defined using the relativistic Hamiltonian (Dirac 1928).

The previous quantization scheme enables the construction of the quantum structure from the underlying classical structure. According to C. J. Isham, "the need for such a tentative approach is rather unsatisfactory, and suggests that the whole idea of 'quantizing' a given classical system is suspect even though, in practice, the procedure does generate many quantum systems of considerable importance … A more logical process would be to start from a quantum system that is given in some intrinsic way, and then to ask about its classical limit" (Isham 1995, 94). The problem is that "there is no clear understanding of what it means to specify a quantum system in an 'intrinsic way'" (Isham 1995, 94). This situation does not prevent taking the quantum theories to be broader than their classical counterparts that are supposed to be recovered from them by some sort of limiting procedure, i.e. it is usually considered that classicality emerges from the quantum realm (see e.g. Landsman 2006, 38).

---

[7] For a nice presentation of the Davisson and Germer experiment see the textbook by Tomonaga (1962, Vol. 2, 10). This ambiguity in the interpretation of the Dirac equation can be seen already in the Schrödinger equation. The one-electron Schrödinger equation can be taken to be a 'classical' equation for a de Broglie wave (see e.g. Tomonaga 1962, Vol. 2, Ch. 6; Becker 1964, Vol 2, 92-107). In Jordan's approach the Schrödinger (or Dirac's) equation was taken to be, implicitly, a 'classical' equation for a matter field, since for Jordan "the 'second quantization' that Dirac had introduced was to be viewed as the quantization of a classical field" (Schweber 1994, 33).

A definition of a limiting procedure in which classical theory appears as some sort of limit of quantum theory presents mathematical and conceptual problems, which have not received an unequivocal answer (Ballentine 1998, 388; see also Landsman 2006). In general the idea of a classical limit is that we might define a sort of mathematical limit that corresponds to a succession of quantum mechanical theories that would take us from quantum to classical physics (Rowe 1991, 1111). In this way "an explicit algorithm may then be used to construct the classical phase space, define a consistent Poisson bracket, and find a classical Hamiltonian, such that the resulting classical dynamics agrees with the limiting form of the original quantum dynamics." (Yaffe 1982, 408) The point is that the 'original' quantum dynamics is constructed by a quantization procedure from the classical description, as was done by Dirac using classical Hamiltonians and the Poisson brackets (see e.g. Kragh 1990, 19). To develop a sort of mathematical procedure to go the other way around can be seen as a consistency check, independently of the interpretation of this procedure as an emergency of classicality.[8] The point of all this is that there is an ambiguity in the interpretation of this procedure. For some it can be seen simply as a consistency check and for others as a limiting procedure. It is only after the argument being presented next (about limitations in the temporal description of physical processes) that I have an independent argument to claim that the so-called limiting procedure is just a consistency check.

To present classical electrodynamics as a limit of quantum electrodynamics is a tricky business, but with the usual long mathematical manipulations we might recover from quantum electrodynamics expressions that look like some expressions of classical electrodynamics, and with this have the impression of obtaining a classical limit of quantum electrodynamics. The point is that in the journey through this mathematical jungle we are loosing sight of the physical interpretation of the formalism of quantum electrodynamics – we only recover some uninterpreted mathematical expressions that resemble expressions from classical electrodynamics but which have not been given a physical interpretation according to the quantum theory they belong to. We can see examples of this, for example, in the work by Stehle and DeBaryshe (1966) and Dente (1975). In the case of Dente the classical limit is supposed to emerge, in the simple case of a particle interacting with an oscillator, by considering the expression for the transition amplitude where the photon coordinates are 'eliminated'. According to Dente the expression for the transition amplitude, "is just the expression which should arise in a classical electrodynamical calculation" (Dente 1975, 1737). The problem is that this is a quantum-electrodynamical expression and should be interpreted accordingly; also this result can be 'read' as a consistency check, i.e. (1) we start from classical physics, (2) we implement a quantization procedure, and (3) we present a 'mathematical' procedure to 'recover' classical-like expressions.[9]

Regarding Stehle and DeBaryshe they call attention to the fact that even the expected 'correspondence' under particular circumstances (according to the authors the high-intensity limit) between quantum-electrodynamical and classical-electrodynamical calculations faces difficulties. They consider in particular the case of the scattering of light by light that has no classical counterpart. According to Stehle and DeBaryshe, "it is not clear that even a low-frequency limit will lead to Maxwell theory" (Stehle and DeBaryshe 1966, 1136). However in the case of the scattering of light by free electrons it is possible to obtain from the quantum-electrodynamical calculation a cross-section equivalent to the classical one (see also Schweber 1961, 638-40). This result does not have to be seen as a mathematical derivation of the classical limit. It can be seen as a consequence of the correspondence principle, which entails, in particular circumstances, 'identical' predictions from both theories (for details on the correspondence principle see e.g. Bokulich 2009; Darrigol 1997).

---

[8] I thank Henrik Zinkernagel for calling my attention to this possibility. This view is also presented by Bohm (1951, 626).

[9] That we should be careful with this type of naïve identification of a 'classical limit' can be seen already in the case of quantum mechanics; see e.g. Ballentine, Yang, and Zibin (1994).

That the so-called 'limiting procedure' does not deliver what it promises can be seen by considering the temporal description of physical processes within quantum electrodynamics as compared to the description we have at the level of classical electrodynamics. In fact, if classical electrodynamics could be seen as a sort of 'mathematical limit' of quantum electrodynamics, the description of processes in the classical theory should also be in some way a limit of an 'underlying' quantum description of the processes. As I will show next, that is not the case.

*3.2 The overall temporal description of physical processes in quantum electrodynamics*

The retardation due to the finite velocity of propagation of the electromagnetic interaction should also be revealed in a quantum electrodynamical treatment by looking at the quantum description of the electromagnetic interaction between charged particles. This would guarantee that it is in principle possible to associate a finite time interval to a particular physical process described at the quantum level. The problem is that quantum electrodynamics does not provide an '*in* time' description of scattering processes (i.e. we cannot associate a finite time interval corresponding to the retardation in the interaction to a physical process like an electron-electron scattering); on the contrary the theory only provides transition probabilities which correspond to measurable relative frequencies, and, as emphasized by B. Falkenburg, "it treats the scattering itself as a black box." (Falkenburg 2007, 131)

The system constituted by the Dirac and Maxwell fields in interaction is described by a set of coupled non-linear equations. It is not possible to find exact solutions of these field equations and the use of perturbative calculations becomes mandatory to the point that, according to F. Dyson, quantum electrodynamics "is in its nature a perturbation theory starting from the non-interacting fields as an unperturbed system." (Dyson 1952, 79) To be able to use free fields in the description of scattering processes two things are necessary: (1) to consider that the initial and final states of the system are eigenstates of the Hamiltonian for the non-interacting fields (intuitively meaning that since the particles are far apart their interaction is negligible), and (2) that the interaction occurs in a short (undefined) period of time when compared to the time it takes the particles to arrive at the 'interaction region' and their posterior observation (which is taken to occur much later when the particles are already not interacting anymore).

This rather vague picture of a scattering process is given a more formal structure within the theory by considering that the interaction between particles in a scattering process is adiabatically switched on from the remote past and adiabatically switched off into the remote future (Lippmann and Schwinger 1950, 473; see also Bogoliubov and Shirkov 1959, 197). This is achieved by considering an initial state at $t = -\infty$ corresponding to free particles and a final state at $t = +\infty$ also corresponding to free particles. This will imply that in this approach, when calculating the so-called S-matrix, we are always considering integrations between $-\infty$ and $+\infty$ in the temporal variable (the same occurs with the spatial variables), while disregarding the detailed description of the intervening times. In this sense we have an *overall* temporal description of the scattering processes, but there really is no description *in time* of the interaction (we are unable to associate a finite time interval to a physical process). From an experimental point of view, things make sense as they are because in scattering processes we only have experimental access to cross-sections. In quantum electrodynamics the scattering cross-section is calculated from the transition probability per unit space-time volume, which is related to the S-matrix in a simple way (see e.g. Jauch and Rohrlich 1976, 163-7). According to Falkenburg, "the effective cross-section is the physical magnitude with which the current field theories come down to earth. As a *theoretical quantity*, the cross-section is calculated from the S-matrix of quantum mechanics … as an *empirical quantity*, it is the measured relative frequency of scattering events of a given type." (Falkenburg 2007, 107)

We might think that this 'black box' description results from a particular method and is not a general feature of the theory; but the case is that, even if this formalism is particularly tailor-made for the description of scattering processes, it is also applicable to bound-state problems (see e.g. Veltman 1994, 62-7). That is, all the calculations made within quantum electrodynamics, giving an excellent agreement with experimental results, can be seen as S-matrix type perturbative calculations (see e.g. Berestetskii, Lifshitz and Pitaevskii 1982, 554).

*3.3 A tentative quantum electrodynamical model providing a temporal description of physical processes*

Even if the theory shows severe limitations if we try to assign a temporal duration to physical processes, we might hope that there are still some other applications of quantum electrodynamics approaching more classical situations where this 'black box' type of description (characterized by an overall space-time approach) does not arise. At first sight, it could seem that a simple treatment of the interaction between two atoms might provide just that: when an atom initially in an excited state decays emitting a photon, it will only be absorbed by a second atom (initially in ground state) after roughly the time r/c (where r is the distance between the atoms and c is the velocity of light), in this way being possible to associate a clear (finite) time interval to the interaction process (i.e. we do not have an overall temporal description of the physical process but an *in time* description like in the classical case).[10]

Suppose that the emission from one atom takes place at a certain region $V_y$ of space-time, meaning that the atom is in a certain location of space and that a photon with energy $\omega_0$ is emitted during a period of time T. This photon is absorbed in the region $V_x$ by another atom. We are going to take for granted for the moment that if the emission takes a time T then there will be an uncertainty $\Delta\omega$ in the energy $\omega_0$ of the photon, so that we have $T\Delta\omega > 1$. At the same time we will consider only a situation where $\Delta\omega$ is much smaller that $\omega_0$, which means that the sign of the energy is defined, and so it is clear that the energy is flowing out of the atom in $V_y$. We thus suppose we can adjust T so that $\Delta\omega \ll \omega_0$. From all this we have that $\omega_0 T \gg 1$. With this condition, considering the part of S-matrix that is due to transitions in $V_x$ and $V_y$ and by taking into account *heuristically* the form of the photon emission intensity distribution and that the atoms are sufficiently apart, every instant of time in the space-time region $V_x$ is greater than every instant of time in $V_y$. The result is that the two bounded regions do not overlap, so that we can say that "if the energy of the charged particles in $V_x$ increases … and if the energy in $V_y$ decreases, then $V_x$ is later in time than $V_y$." (Pauli 1973, 133) Also, the second region must be on the (diffuse) light cone of the first. For this reason, besides an uncertainty ±T resulting from "the uncertainty in the time of the emission process," (Thirring 1958, 146) we only have contributions in the S-matrix from space-time points in regions that can be connected by photons propagating at light speed. This also means that for the distance r between the two atoms we have $r\omega_0 \gg 1$: the second atom must be in the wave zone of the first (roughly speaking a region where r is much bigger than the wavelength of the emitted photon: $r \gg \lambda$).

---

[10] I will be considering Fierz's (1950) S-matrix approach to the interaction between two bound electrons. This approach is presented in the textbooks by Pauli (1973) and Thirring (1958).

Apparently everything is just fine and we have a model of the interaction of two bound electrons in which we obtain a clear temporal relation between the emission and absorption of a photon corresponding to a retarded interaction as we expect from classical electrodynamics. However a closer look into the development of the model shows that this is not the case.

Maybe the most interesting aspect of this model is that in the wave zone we see that the contribution from the photon propagator in the S-matrix comes from the poles, corresponding to a process with a real photon (Thirring 1958, 146). As Feynman remarked, "in a sense every real photon is actually virtual if one looks over sufficiently long times scales. It is always absorbed somewhere in the universe. What characterizes a real photon is that $k \to 0$." (Feynman 1962, 95) *We see that the distinction between virtual and real photon can depend on the separation between the atoms: in the near zone ($r \ll \lambda$) the photons are virtual and in the wave zone they are real*.[11] One point is clear from the previous example. To approach an idea of temporality (i.e. temporal order in the physical processes) in models of interaction, using as a fundamental part quantum electrodynamics, we need structures in the models – the atoms – that permit the appearance of real photons, which approach a more classical electrodynamical type of interaction (emission and posterior absorption of light). Also for a consistent outcome from this model it is necessary that $\omega_0 T \gg 1$ and this is not provided by the theory directly, because there is no time-energy uncertainty relation in quantum theories in the same sense as in non-relativistic quantum mechanics where there is an uncertainty relation between the position and momentum operators (Hilgevoord 1996, 1451). We have somehow to go and get it.

In the Maxwell-Lorentz classical theory we have a relation between $\Delta\omega$, the line width, and the lifetime T of the radiation emission process. Considering a linear harmonic oscillator (as a model for a light source) and taking into account the effect of the field produced by the charge on the charge itself (the self-force), the (emission) intensity distribution is given by

$$I(\nu) = I_0 \frac{\gamma}{2\pi} \frac{1}{(\nu - \nu_0)^2 + \gamma^2/4},$$

---

[11] In the space-time points where a virtual photon is created or annihilated we have conservation of energy and momentum between the photon and the electrons. However the energy-momentum relation for the virtual photon is not $k^2 = (k^0)^2 - \mathbf{k}^2 = 0$ corresponding to a zero mass real photon, it is different from zero due to the fact that in the expression for the propagator of the photon (which is part of the S-matrix) $\mathbf{k}$ and $k^0$ are independent of each other (e.g. Mandl and Shaw 1984, 86), as if the virtual photon has a mass during the interaction process. Also the propagator of a virtual photon is nonvanishing at space-like separations (e.g. Björken and Drell 1965, 388-9).

where $\nu_0$ is the frequency of the undamped oscillator, and $\gamma$ is the width at half the maximum intensity and is equal to the reciprocal of the lifetime of the oscillator (due to the damping of the self-force the oscillator radiates during a period of time until it comes to rest) (see e.g. Heitler 1954). Under the assumption that the reaction force is small we have that the lifetime is long when compared with the period of the oscillator, so that we have $\gamma \ll \nu_0$, that is, $\omega_0 T \gg 1$. Following the same approach when describing the decay of an excited state of an atom in quantum theory, again it is assumed that "the lifetime is large compared with the frequency of the atom," (Heitler 1954, 183) that is $\omega_0 T \gg 1$, and we obtain the same expression for the intensity distribution of the emission (now as a probability function). In this way, we do not use any 'uncertainty' relation to obtain the result $\omega_0 T \gg 1$ which is needed to obtain a model for the interaction between the atoms that appears to give a consistent spatio-temporal description of the interaction. We see then that the model is dependent on results that can also be derived in a classical treatment of the emission of radiation by an atom.

By a *heuristic procedure* (and not by a first-principles derivation), taking into account the relation $\omega_0 T \gg 1$ and the corresponding emission line, a specific form is given to the bilinear density $\overline{\psi}\gamma^\mu\psi$ in the second-order term of the S-matrix used in this model:

$$\text{const} \int_{V_x} d^4x \int_{V_y} d^4y (\overline{\psi}(x)\gamma^\mu\psi(x)) D_c(x-y)(\overline{\psi}(y)\gamma^\mu\psi(y)),$$

where $D_c(x-y)$ is the photon propagator, with $\overline{\psi}\gamma^\mu\psi \sim a_1 a_2^* \rho_\mu(x) \exp[i\omega_0 t - t^2/T^2]$, and $\omega_0 > 0$.[12] It is from this expression that the desired temporal behaviour is obtained.

We have then in the 'core' expression of the model a term defined at the correspondence limit, i.e., a term that can be derived either by classical electrodynamics or the quantum theory. In this way the classically derivable spectral line curve is a fundamental aspect of this model.[13] In simple terms what I mean by 'correspondence limit' is the asymptotic agreement between the predictions of quantum and classical mechanics. According to Bohr the correspondence principle can be seen as a law of quantum theory (see e.g. Bokulich 2009). His view is that "the correspondence principle expresses the tendency to utilise in the systematic development of the quantum theory every feature of the classical theories in a rational transcription appropriate to the fundamental contrast between the [quantum] postulates and the classical theories" (quoted in Bokulich 2009, 18). In Heisenberg's and Bohr's view matrix mechanics embodies the correspondence ideas by a 'symbolic' translation of the Fourier components of the motion to *corresponding* quantum amplitudes. According to O. Darrigol in this last formulation of the 'correspondence principle' as part of quantum theory we have a 'symbolic translation' of classical concepts (i.e. we obtain the corresponding quantum concepts by the quantization procedure) and maintain the (statistical) asymptotic agreement between the quantum and the classical theories (Darrigol 1997). It is important to notice that this asymptotic agreement (correspondence limit) is inbuilt in the theory and it is not related to any eventual 'mathematical limiting procedure' that recovers the classical theory from the quantum theory.

---

[12] We see that a Gaussian function is being used. The expected function according to classical theory (or its quantum counterpart) is a Lorentzian function. The Gaussian can be obtained when considering the broadening due to the Doppler effect occurring in a 'gas' of atoms at temperature T (e.g. Major 1998, 142-9). That is, the expression being used is not valid in the case of a single atomic system, which is the case being considered in this model.

[13] This point is not made in the presentation of the model and its results (Fierz 1950, 734-5; Pauli 1973, 134-5).

The implication of all this is that the model can be seen as a semi-classical one. Thus when using this model we are at the 'surface' of contact of the classical and quantum theories – we are not 'diving' into the quantum world, which is supposed to be underlying the classical mode of description. And what we are looking for is an (*in time*) temporal description (if any) of physical processes in this 'quantum world', i.e. the description given in a full quantum electrodynamical calculation.

The final ingredient in the development of the model is an adjustment by hand of the distance between the atoms so that the second atom lies in the wave zone of the first. As mentioned, in the near zone the photon behaves as a virtual one, while in the wave zone we have (as a limit) the energy-momentum relation for a 'real' photon, which means that in the last leg of the model development, depending on how we choose the distance between the atoms, we can have a situation where we can associate a causal temporal order to the emission and absorption process of a 'real' photon, or a situation where it is not possible to associate a causal temporal order to the emission and absorption of a 'virtual' photon. In this way *we choose* the photon to be real, and we find that it is possible to associate a causal temporal order to the emission and absorption process of the 'real' photon.

Another difference between this model and the S-matrix calculations of scattering (or bound state) processes is that in the second case we obtain results that can be compared to experiments (see e.g. Falkenburg 2007, 105-7), while in the first case we can only associate with the model a thought-experiment (Buchholz and Yngvason 1994, 613). In this way the model patched from the theory gives the impression of a solid verifiable consequence that it really is not.

To sum up: we are using a model that describes a thought-experiment; ultimately the main input that determines the form of the bilinear density in the S-matrix is not provided by a Dirac equation computation but by a heuristic use of what can be seen as a classical result for the radiation emission of a bound electron; and finally, an adjustment 'by hand' of the distance between the atoms is made, to make it possible to obtain the desired temporal behaviour. This does not seem to be a solid procedure we can use to defend that we can retain within a quantum treatment the possibility of the kind of temporal description of processes that we have in classical electrodynamics.

In fact, this model cannot be seen as an application of the quantum electrodynamical formalism. More properly it is a heuristic semi-classical model, which, as mentioned, 'works' at the correspondence level of description (since part of its elements are described by classical physics). I think that to arrive at any conclusive results, we have always to stick to the physical interpretation of quantum theories and consider clear applications of quantum electrodynamics related with doable experiments. Here we have seen that the theory only provides 'black box' descriptions of physical processes, corresponding to an overall space-time approach. In this way from what has been presented *we cannot consider that from the quantum level of description it is possible to recover the temporal description of processes that we have at a classical level*.

*Conclusion*

Considering the results presented here, there seems to be no smooth and physically unproblematic way to connect quantum electrodynamics with classical electrodynamics.

Thus the status of quantum electrodynamics has to be reviewed. Quantum electrodynamics cannot be seen as an independent theory that "contains [classical] electrodynamics as a special case." (Stehle and DeBarysche 1966, 1135) It seems more adequate to regard quantum electrodynamics as a physical-mathematical upgrade of classical theory (electrodynamics and the theory of relativity), which permits an extension of the range of application of the classical theory in the description of natural phenomena, but which does not operate a reduction of the classical theory. This conception of quantum electrodynamics as something constructed on top of classical physics fits well with O. Darrigol's idea of a modular structure of physical theories.[14] Thus quantum electrodynamics can be seen as part of a broader theoretical modular structure that is expected to describe –with both its classical and quantum parts– what we consider to be the phenomena of matter, radiation and their interaction. Quantum electrodynamics works as an extension (or upgrade module) of the classical theory into 'regions' where this fails completely, but since it has been developed from classical concepts, and its probabilistic interpretation puts clear constraints on the applicability of the theory, we cannot expect to recover the classical part of the description of the phenomena from the quantum part by some limiting procedure. In particular, it is not possible to recover fully, from quantum electrodynamics, the kind of temporal description of physical processes that we have with the classical theory.

---

[14] According to Darrigol, "the practice and the history of physics show that physical theories are not homogenous wholes, and that a given physical theory is usually used in conjunction with other theories … I introduce the notion that any non-trivial theory has essential components, or *modules*, which are themselves theories with different domains of application" (Darrigol 2008, 195-6).